\documentclass[10pt,twocolumn,letterpaper]{article}

\usepackage[pagenumbers]{wacv} 
\usepackage{graphicx}
\usepackage{amsmath}
\usepackage{amssymb}
\usepackage{booktabs}
\usepackage{float}
\usepackage{caption}
\usepackage{multicol}

\usepackage[pagebackref,breaklinks,colorlinks]{hyperref}

\usepackage[capitalize]{cleveref}
\crefname{section}{Sec.}{Secs.}
\Crefname{section}{Section}{Sections}
\Crefname{table}{Table}{Tables}
\crefname{table}{Tab.}{Tabs.}

\def\wacvPaperID{*****} 
\def\confName{WACV}
\def\confYear{2024}

\begin{document}

\title{Decoding visual brain representations from electroencephalography through  Knowledge Distillation and latent diffusion models}

\author{Matteo Ferrante\\
Department of Biomedicine and Prevention, \\ University of Rome Tor Vergata (IT)\\
{\tt\small matteo.ferrante@uniroma2.it} \\
\and
Tommaso Boccato\\
Department of Biomedicine and Prevention, \\ University of Rome Tor Vergata (IT)\\
\and 
Stefano Bargione \\
Department of Biomedicine and Prevention,\\ University of Rome Tor Vergata (IT) \\
\and 
Nicola Toschi \\
Department of Biomedicine and Prevention,\\ University of Rome Tor Vergata (IT) \\
Martinos Center For Biomedical Imaging, MGH and Harvard Medical School (US)
}
\maketitle

\begin{abstract}

Decoding visual representations from human brain activity has emerged as a thriving research domain, particularly in the context of brain-computer interfaces. Our study presents an innovative method that employs to classify and reconstruct images from the ImageNet dataset using electroencephalography (EEG) data from subjects that had viewed the images themselves (i.e. "brain decoding"). We analyzed EEG recordings from 6 participants, each exposed to 50 images spanning 40 unique semantic categories. These EEG readings were converted into spectrograms, which were then used to train a convolutional neural network (CNN), integrated with a knowledge distillation procedure based on a pre-trained Contrastive Language-Image Pre-Training (CLIP)-based image classification teacher network. This strategy allowed our model to attain a top-5 accuracy of 80\%, significantly outperforming a standard CNN and various RNN-based benchmarks. Additionally, we incorporated an image reconstruction mechanism based on pre-trained latent diffusion models, which allowed us to generate an estimate of the images which had elicited  EEG activity. Therefore, our architecture not only decodes images from neural activity but also offers a credible image reconstruction from EEG only, paving the way for e.g. swift, individualized feedback experiments. Our research represents a significant step forward in connecting neural signals with visual cognition.
\end{abstract}


\section{Introduction}

\begin{figure*}[t]
    \centering
    \includegraphics[width=1\linewidth]{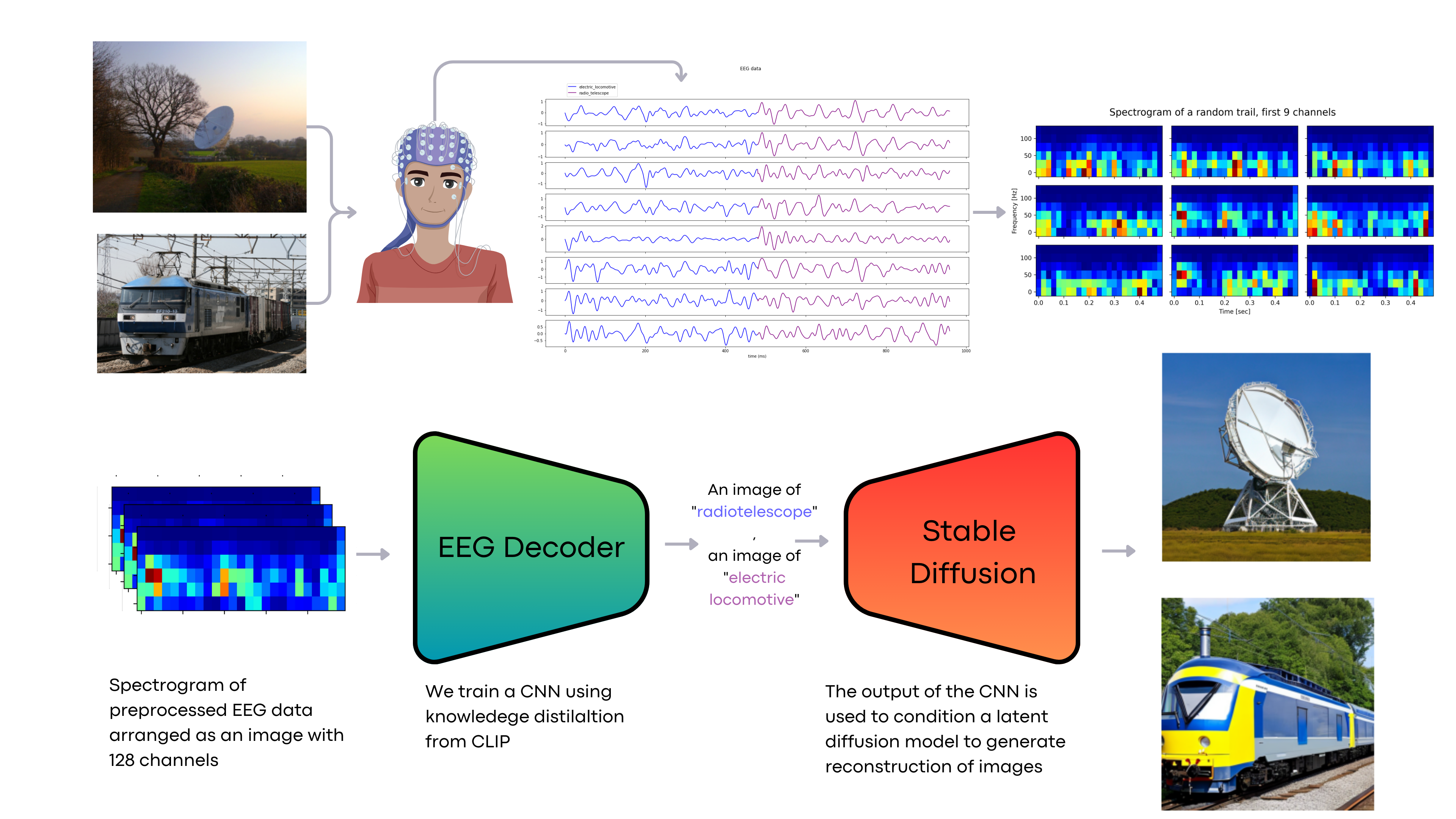}
    \caption{Our pipeline can be described as follows: First, we record EEG data while the subject is viewing natural images. This data is then preprocessed and converted into spectrograms, which serve as the input for our neural network. Our EEG decoder is trained using a knowledge distillation method based on the CLIP model. The outputs from the EEG decoder, which are predictions of the image that elicited the EEG data, are then combined with an image generation pipeline. This end-to-end approach allows us to reconstruct images from the neural activity data captured by the EEG.}
    \label{fig:scheme}
    \vspace{-5mm}
\end{figure*}

Electroencephalography (EEG) is increasingly recognized as a valuable instrument for decoding visual representations within the human brain. The  primary advantage of EEG  lies in its non-invasive nature and its ability to provide real-time insights into human brain function via electrical activity recordings from the scalp. Despite its spatial resolution constraints, its unparalleled temporal resolution renders it ideal for real-time applications. 

Recent technological advancements have facilitated the decoding of intricate visual stimuli from EEG signals, notably from expansive datasets such as ImageNet \cite{Palazzo,bai2023dreamdiffusion}. Both convolutional (CNN) and recurrent neural networks (RNN) have demonstrated efficacy in classifying EEG signals into distinct image categories with appreciable accuracy.  The successful decoding of complex visual stimuli from EEG signals can pave the way for innovative neural prosthetics and biofeedback systems. Translating brain activity patterns into decoded image categories or reconstructions could potentially offer visually impaired individuals a semblance of artificial vision. Additionally, EEG decoding can revolutionize brain-centric image searches, communication platforms, and augmented reality interfaces. Real-time visualizations of decoded brain activity can also usher in novel neurofeedback paradigms, facilitating self-regulation of brain states through integrated EEG decoding and external visual feedback mechanisms \cite{eeg_feedback}.

However, a predominant focus in current research is on multisubject models, which involve averaging EEG signals across multiple participants. This methodology may overlook the nuances of individual-specific neural representations. Models tailored to individual subjects could offer a more granular decoding and introduce an added dimension of data privacy, as each model is uniquely calibrated for a specific individual, precluding its application to others. Also, in spite of recent progress, the task of reconstructing visual stimuli based on the  EEG signals they elicit remains a formidable challenge. The inherent low spatial resolution of EEG poses difficulties in reconstructing detailed visual nuances. Presently, image reconstructions predominantly capture broader features, such as shapes, colors, and textures, thereby constraining the depth of visual feature decoding and image reconstructions. To overcome this obstacle, instead of attempting pixel-precise reproductions, a more pragmatic approach might be semantic image reconstructions. In this context, techniques such as generative adversarial networks (GANs) Here's the revised LaTeX text with a more sober and scientific style:

The emergence of methods such as \cite{Brain2Image} offers semantically coherent reconstructions directly from EEG signals, rather than estimating the EEG from an image and subsequently attempting to reconstruct the image from this estimated signal. Despite the challenges associated with the fidelity of image reconstruction, the rapid advancements in deep neural networks show potential. This research aims to improve existing methodologies for translating perceptual experiences from EEG patterns, with a focus on real-time applications. We present a methodology that advances this field, outlining a pipeline (as shown in Fig \ref{fig:scheme}) that facilitates the training of a single-subject model within a limited experimental timeframe, leading to near-real-time brain decoding.

\section{Related Works}
EEG are widely processed in the context of brain-computer interfaces (BCI) to perform brain decoding for a wide variety of tasks \cite{zafar_decoding_2015}.
A number of prior works have explored decoding visual representations from EEG signals using deep learning models. Kavasidis et al. \cite{kavasidis_brain2image_2017} were among the first to propose generating images from EEG data. They recorded EEG while subjects viewed ImageNet images, and used an Long Short Term Memory (LSTM) model combined with variational autoencoders or GANs to reconstruct images. The key difference is they aimed for class-level image generation rather than detailed reconstruction and focuses on processing data in the time domain.
Spampinato et al. \cite{spampinato_deep_2017} also analyzed EEG responses to ImageNet stimuli. They trained an LSTM encoder to classify EEG signals into image categories. For reconstruction, they trained a separate CNN regressor to predict EEG features from images and replaced the EEG signal with this encoder model. 
Palazzo et al. \cite{palazzo_generative_2017} extended \cite{spampinato_deep_2017} using contrastive learning to align EEG and visual image features. However, their goal was improving image classification rather than reconstruction, and various challenges emerged  \cite{trainingtest}. Singh et al. \cite{singh_eeg2image_2023} proposed an EEG-to-image GAN framework but focused on smaller (i.e. with fewer images) datasets of characters and shapes.
In this work, we propose a modularized pipeline for reconstructing detailed photorealistic visual stimuli (i.e. images) directly from EEG brain signals, using a novel CLIP based knowledge distillation of a convolutional neural network trained on time-frequency decomposition (TFD) and generative diffusion synthesis, generating semantically plausible and visually similar images reconstruction to the original stimuli.

\section{Material and Methods}

This section delineates the methodology adopted and the dataset utilized. The dataset, sourced from ImageNet EEG \cite{Brain2Image}, is publicly accessible. All computational experiments and model training were conducted on a server outfitted with four NVIDIA A100 GPU cards (each with 80GB RAM connected via NVLINK) and 2 TB of system RAM. The codebase was developed using Python 3.9, leveraging libraries such as Pytorch, Pytorch Lightning, and scikit-learn for model implementation.

\subsection{Data}

The EEG recordings employed in this study were sourced from \cite{spampinato2019deep}. These recordings were obtained from six subjects who were exposed to images from 40 distinct ImageNet \cite{imagenet} classes, with each class comprising 50 images. The sampling rate for these recordings was 1000 Hz. The image presentation protocol involved sequential display in 25-second intervals, succeeded by a 10-second intermission. In each display interval, images are shown sequentially for 0.5 seconds each. This protocol yielded a total of 2,000 images spanning 1,400 seconds (or 23 minutes and 20 seconds) of recording time. Each subject underwent four recording sessions, each lasting 350 seconds. 
The experiments utilized a 128-channel cap with active, low-impedance electrodes (actiCAP 128Ch, Brainproducts) 
for EEG data collection. Brainvision 
amplifiers and data acquisition systems were used to record the EEG signals at a sampling rate of 1000 Hz with 16-bit resolution. The EEG data resulted in 11,466 sequences post the exclusion of recordings of suboptimal quality. 
The comprehensive nature of this experimental design facilitated the examination of EEG responses to a diverse array of visual stimuli from ImageNet. The multi-channel EEG recordings, captured during the viewing of thousands of stimuli, furnish a rich dataset conducive for training decoding models. For further detail about acquisition protocol please see the original article \cite{spampinato2019deep}.

\subsection{Preprocessing}

Prior to utilizing the  EEG signals for training our decoding models, a series of preprocessing steps were executed. Initially, a notch filter in the 49-51 Hz range was applied to mitigate power line interference. Subsequently, a second-order band-pass
Butterworth filter, ranging between 14 and 70 Hz, was employed to focus on frequency bands pertinent to visual attention and object recognition. The signals were then standardized across channels.
For the purpose of neural network input generation, the filtered EEG signals were segmented into 40 ms windows moving each time 20 ms. Time-frequency decompositions (TFD) were computed for these segments using the short-time Fourier transform (STFT), converting each trial into a 128-channel image that depicted the spectrum across both time and frequency dimensions. This process yielded 2,000 EEG spectrogram images, each with 128 channels, for every subject. These images were then used for training and evaluation of our convolutional neural network tailored for EEG decoding. This multi-channel spectral representation encapsulates the spatial and temporal intricacies of the EEG, allowing our model to extract features essential for visual stimuli classification. It is worth noting that the preprocessing described herein is specific to the architecture proposed in this study. Alternative baselines adopted slightly varied preprocessing techniques, such as direct time domain data analysis, starting from the same filtered data in the time domain. These variant preprocessing methodologies are elaborated upon in \ref{sec:Baselines}.

\subsection{Decoding pipeline}

Our approach employs a CNN with integrated residual connections to classify EEG TFDs. The architecture begins with a series of convolutional layers, progressively increasing the number of filters to effectively extract both spatial and temporal features. Subsequent to this, global average pooling and fully-connected layers are utilized for classification tasks.
For the training of the CNN, we adopt a knowledge distillation methodology \cite{hinton2015distilling}. Initially, an image classifier is pretrained using CLIP (Contrastive Language-Image Pre-Training) \cite{radford2021learning} features to anticipate the stimulus classes, achieving a commendable accuracy of 99\%. This pretrained classifier furnishes "soft targets" to guide our EEG model.
During the training phase, EEG spectrograms are fed into the CNN, while CLIP image features are directed to the teacher classifier. The objective is to train the CNN such that it aligns with the class probability distributions produced by the teacher. This distillation approach not only stabilizes the training process but also enhances the model's performance in comparison to direct training on class labels.
For inference, only the EEG-based CNN is deployed to predict classes from novel time-frequency decompositions. Through the distillation of knowledge from the image model, our CNN is equipped to derive robust representations, enabling the decoding of visual stimuli solely from EEG signals.

\begin{figure}
\centering
\includegraphics[width=1.1\linewidth]{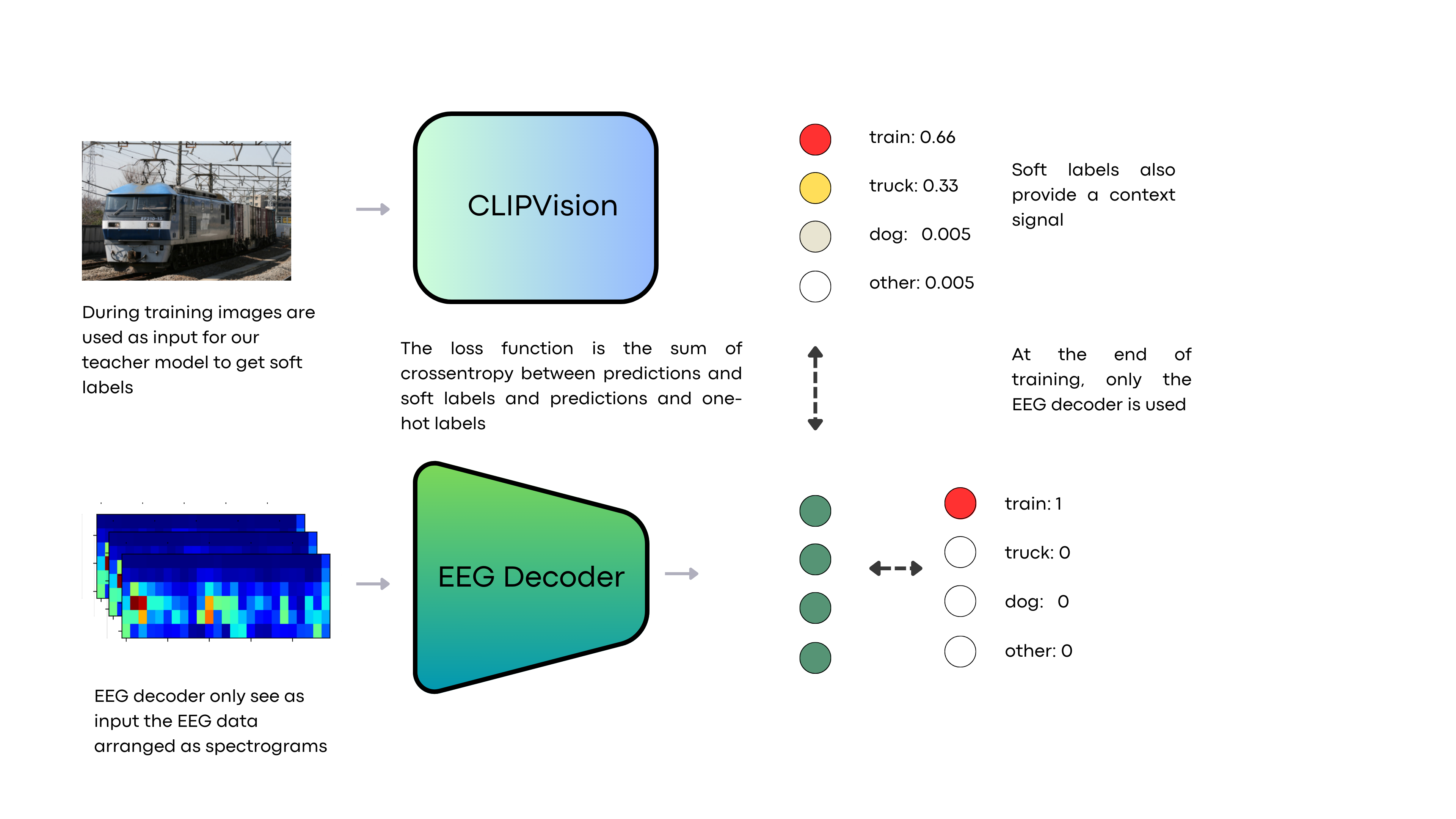}
\caption{Illustration of the training procedure. Knowledge distillation facilitates the training of a compact "student" model to emulate the outputs of a more extensive "teacher" model. This enables the student to achieve performance levels akin to larger models, even when initiated from distinct yet related inputs.}
\label{fig:KD}
\end{figure}

Post the training of our EEG decoding model, it becomes capable of predicting ImageNet classes from fresh EEG TFDs. To validate these predictions and reconstruct images that could potentially induce analogous neural responses, we employ the Stable Diffusion generative model \cite{stable}. For every EEG prediction, a text prompt such as "an image of a {predicted class}" is formulated. This prompt, in conjunction with random noise vectors, is input into Stable Diffusion to generate images congruent with the predicted class. This methodology facilitates the reconstruction of visual stimuli exclusively from neural activity patterns. The EEG decoder identifies the class, while Stable Diffusion fabricates a semantically coherent image. A comprehensive diagram of the decoding pipeline is depicted in Fig \ref{fig:scheme}, and the knowledge distillation procedure is illustrated in Fig. \ref{fig:KD}.

\subsection{Reconstruction Pipeline}

Diffusion models are generative frameworks trained to invert a noise diffusion process, facilitating image synthesis. Stable Diffusion operates as a latent diffusion model, proficient in generating lifelike images from random noise vectors, conditioned by textual descriptions. The model's strategy involves the iterative addition of noise to genuine images, followed by the learning of a parametric denoising function to eradicate the noise over multiple timesteps. By repetitively applying the denoising function, the model can produce lifelike images, conditioned on textual descriptions. This iterative denoising offers tight control over image generation, guided by text at every iteration.
In the sampling phase, Stable Diffusion accepts a text prompt and progressively diffuses noise vectors until they converge into an image that aligns semantically with the provided description.
For the task of reconstructing images from EEG signals, Stable Diffusion's text conditioning capability proves invaluable. The EEG decoder outputs a label indicative of the visual stimulus class. This discrete label is then employed to generate corresponding images via Stable Diffusion, bypassing the need for direct pixel reconstruction.
This approach facilitates the synthesis of plausible image reconstructions based on the decoded semantic category from neural activity patterns. This model-centric strategy also addresses the inherent resolution constraints of EEG for high-fidelity decoding. The guided diffusion modeling ensures the generation of visualizations that are both realistic and interpretable to human observers.

\subsection{Knowledge Distillation}

Knowledge distillation facilitates the transfer of insights from a comprehensive, pretrained teacher model to a more compact student model \cite{hinton2015distilling}. This process empowers the student model to attain performance metrics that are typically associated with larger models.

Consider $f_{t}(x)$ as the output vector of class probabilities produced by the teacher model for a given input $x$, representing the stimulus image. Similarly, let $f_{s}(e;\theta)$ denote the student model, characterized by parameters $\theta$, where $e$ represents the EEG recordings obtained during the presentation of stimulus $x$. The student model is trained through knowledge distillation by minimizing:

\begin{equation}
\mathcal{L}(\theta) = \alpha \mathcal{L}{CE}(f{s}(e;\theta), y) + (1 - \alpha) \mathcal{L}{KD}(f{s}(x;\theta), f_{t}(x))
\end{equation}

Here, $\mathcal{L}{CE}$ represents the cross-entropy loss between the predictions of the student model and the actual ground truth labels $y$. In contrast, $\mathcal{L}{KD}$ denotes the distillation loss, capturing the difference between the outputs of the student and teacher models. The temperature parameter $T$ is employed to modulate the probability distribution of the teacher:

\begin{equation}
\mathcal{L}{KD}(f{s}, f{t}) = -\sum_{c} \frac{\exp(f_{t,c}/T)}{\sum_{c'} \exp(f_{t,c'}/T)} \log \frac{\exp(f_{s,c}/T)}{\sum_{c'} \exp(f_{s,c'}/T)}
\end{equation}

Training the student model to replicate the comprehensive probability distribution of the teacher facilitates the transfer of insights regarding inter-class relationships, offering a richer supervisory signal than mere ground truth labels. In our implementation, we set $\alpha=0.5$ and $T=1$.

For EEG decoding, a linear classifier was trained atop the CLIP \cite{radford2021learning} CLS tokens. CLIP, an acronym for Contrastive Language-Image Pre-Training, is a neural architecture trained to correlate images and text through contrastive learning. Comprising an image encoder and a text encoder, CLIP is trained to discern whether an image-text pairing is congruent or not. The image encoder in CLIP, a vision transformer (VIT), embeds images into latent representations. Throughout its training, CLIP cultivates an embedding space where semantically congruent images and texts are proximate. A pivotal element of the image encoder is the CLS token, an auxiliary token introduced to the network's input, enabling the encoder to generate a holistic representation of the entire image. A linear classifier was trained atop this CLS token for every image in the training dataset to predict the appropriate class. This amalgamation of CLIP and the classifier served as the teacher model, functioning as a bridge between EEG spectrograms and image classes. The student CNN, when exposed solely to EEG data, derives insights from both the teacher's distributions and the true labels. This distillation process accentuates the student's focus on neural patterns pertinent to visual recognition, enhancing convergence, accuracy, and generalization. By assimilating insights from a domain expert in image processing, the streamlined student decoder becomes adept at extracting visual representations from EEG signals.

\subsection{Baselines}

\label{sec:Baselines}

In order to underscore the efficacy of employing computer vision techniques for EEG signal decoding, we assessed a spectrum of baseline methodologies, spanning from conventional machine learning paradigms to contemporary neural network architectures.

Initially, we employed a basic baseline wherein the raw EEG signals were standardized, squared, and subsequently averaged across channels. Following this, a Logistic Regression classifier was trained on the resultant data. An extension of this approach involved applying the Logistic Regression classifier to EEG signals that were averaged over an 80-point sliding window. In another variant we executed PCA on the windowed average EEG, preserving 29 components that accounted for $95\%$ of the variance, prior to classifier training. Notably, these methodologies overlook the inherent spatial and temporal intricacies of the EEG signal. The main advantage of using the PCA is providing orthogonal features to the model that already integrate relevant spatiotemporal relationships.

In this context, a recent proposition by CEBRA \cite{cebra} demonstrated a deep learning technique that employs contrastive learning to project neural data onto lower-dimensional manifolds conducive for decoding. In alignment with this, we projected our EEG data onto a 32-dimensional manifold, utilizing CLIP features as a guiding mechanism. The value was chosen to be close to the number of features used in the PCA, picking the closest power of 2. This offers a robust nonlinear neural baseline that effectively harnesses both spatial and temporal patterns.

In terms of neural network architectures that directly process EEG time series data, we examined both a LSTM model and a 1D convolutional network (CNN) equipped with temporal convolutions. Both architectures incorporated 4 layers and were regularized using dropout, ensuring a consistent parameter count across models.

Further, we explored CNNs that operate on 2D representations of the EEG, thereby leveraging computer vision methodologies. One such model treated the raw EEG traces as a 2D image. Another model employed a wavelet decomposition utilizing the Daubechies db4 wavelet from PyWavelets [2] \cite{pywavelet}, which has been recognized as an efficient time-frequency representation for EEG \cite{wavelet}. Our final CNN baseline ingested the short-time Fourier transform (STFT) of the EEG, processed with a 40 ms window.

This ensemble of baselines, ranging from classical signal processing to avant-garde deep learning, offers a holistic comparative framework and accentuates the significance of spatiotemporal neural network modeling in the realm of EEG decoding. The computer vision-oriented strategies adeptly harness the structural nuances present in the multi-channel EEG.

For consistency, all neural networks were evaluated with a similar parameter count range (1.1-1.2 M). Each was trained using the Adam optimizer at a learning rate of $3e-4$. Additional training specifications included an early stopping callback with a 10-epoch patience based on validation loss variations, a batch size of 64, gradient clipping at a magnitude of $1.0$, and a maximum epoch count set to 50.

\begin{figure*}[h]
\centering
\includegraphics[width=0.9\linewidth]{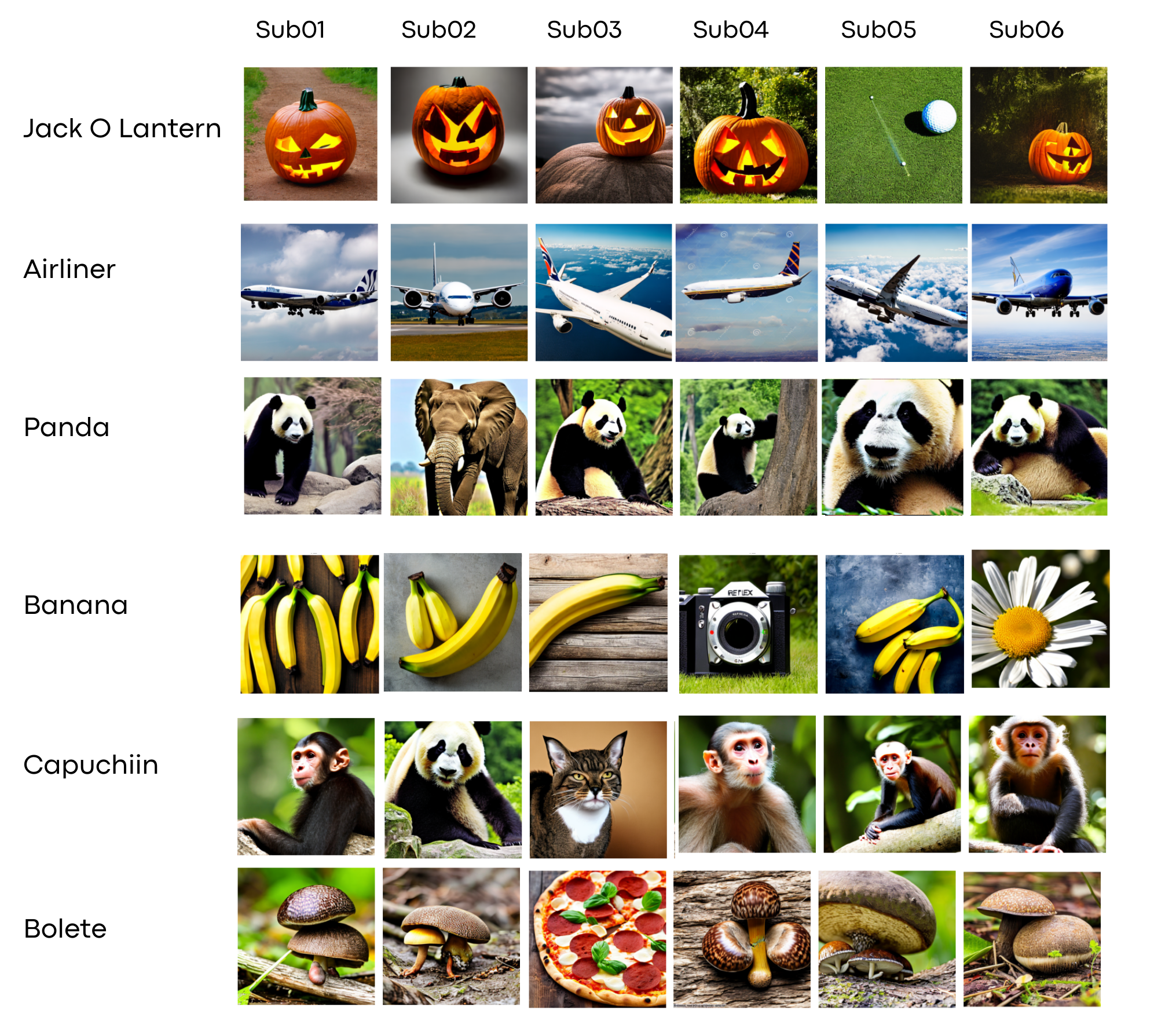}
\caption{Reconstructed images. Left column: target classes; subsequent columns: results from individual subjects.}
\label{fig:qualitative}
\end{figure*}

\section{Results}

\begin{figure}[t]
    \includegraphics[width=1.\linewidth]{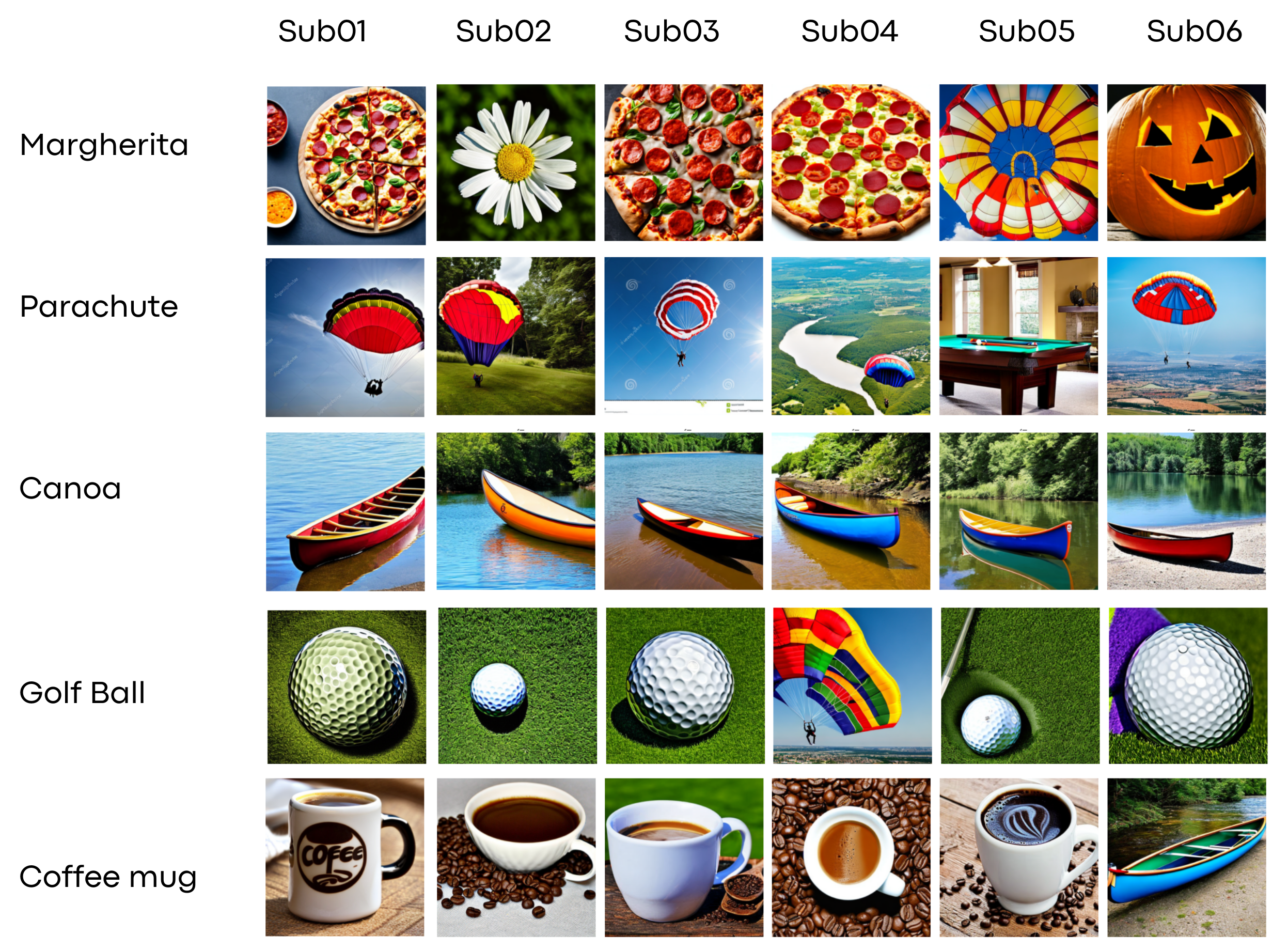}
    \caption{On the left, the target classes are presented and each column show result from a single subject.}
    \label{fig:other_qualitative}
\end{figure}

\begin{table*}
  \centering
  

  \begin{multicols}{2} 

    \resizebox{\linewidth}{!}{\begin{tabular}{@{}lcccccc@{}}
      \textbf{Method} & \multicolumn{6}{c}{\textbf{Metrics [Mean (Std)]}} \\
             & \textbf{Accuracy} & \textbf{Top3 Accuracy} & \textbf{Top5 Accuracy} & \textbf{F1} & \textbf{Kappa} \\
      LR on average square signal & 0.3600 (0.1313) & 0.6619 (0.1758) & 0.8156 (0.1619) & 0.3493 (0.1375) & 0.3435 (0.1345) \\
      LR on windowed signal & 0.0205 (0.0058) & 0.0636 (0.0083) & 0.1092 (0.0110) & 0.0156 (0.0054) & 0.0009 (0.0061) \\
      LR on PCA windowed signal & 0.0175 (0.0040) & 0.0536 (0.0084) & 0.0961 (0.0063) & 0.0097 (0.0047) & 0.0020 (0.0039) \\
      CEBRA + kNN & 0.0240 (0.0050) & 0.0831 (0.0116) & 0.1402 (0.0136) & 0.0223 (0.0061) & -0.0012 (0.0056) \\
      LSTM & 0.3605 (0.0938) & 0.7376 (0.1226) & 0.8868 (0.1030) & 0.3392 (0.0894) & 0.3437 (0.0960) \\
      Conv1d & 0.2623 (0.0511) & 0.6013 (0.0826) & 0.7971 (0.0851) & 0.2582 (0.0520) & 0.2432 (0.0524) \\
      Knowledge distillation on eeg (img) & 0.2819 (0.0836) & 0.5773 (0.1379) & 0.7295 (0.1339) & 0.2742 (0.0794) & 0.2632 (0.0857) \\
      Knowledge distillation on wavelet & 0.4060 (0.1154) & 0.7490 (0.1282) & 0.8787 (0.1007) & 0.3889 (0.1148) & 0.3905 (0.1183) \\
      plain CNN on spectrograms & 0.2819 (0.0836) & 0.5773 (0.1379) & 0.7295 (0.1339) & 0.2742 (0.0794) & 0.2632 (0.0857) \\
      \textbf{Knowledge distillation on STFT} & \textbf{0.4120 (0.1131)} & \textbf{0.7530 (0.1068)} & \textbf{0.8782 (0.0806)} & \textbf{0.4027 (0.1133)} & \textbf{0.3966 (0.1160)} \\
    \end{tabular}}

  \label{tab:example}
  \end{multicols} 
\caption{Performance comparison of various methods applied to EEG signal data. The table presents the mean values accompanied by the standard deviation (enclosed in parentheses) for each evaluation metric across all subjects. Notably, the knowledge distillation techniques, when applied TFD generated using STFT and wavelet decompositions, yield the best results.}
\end{table*}

\subsection{Performance Evaluation}

\begin{figure}[t]
    \centering
    \includegraphics[width=.9\linewidth]{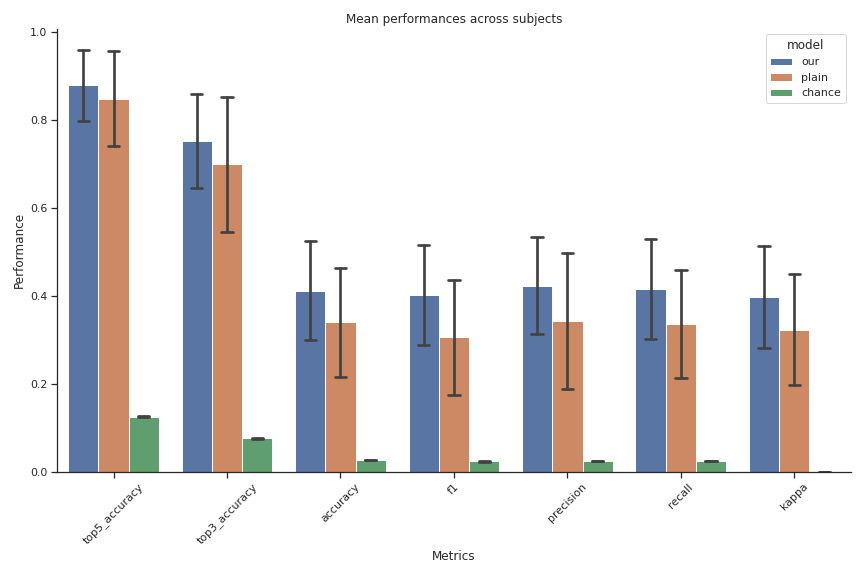}
    \caption{Resuls for EEG decoder. \textbf{Ours} is the CLIP-based approach, \textbf{plain} is a vanilla CNN with the same architecture trained for classification and \textbf{chance} serves as comparison with chance level. Bars are average across subjects and error bars are standard deviations.}
    \label{fig:metrics}
\end{figure}

The efficacy of our model is evaluated using a comprehensive set of metrics: top-5, top-3, top-1 accuracy, F1 score, and the normalized kappa score. Figure \ref{fig:metrics} demonstrates that our knowledge distillation CNN consistently outperforms both the standard CNN baseline and a random classifier. Notably, the proposed approach—employing a CNN on TFD with CLIP-based knowledge distillation—exhibits superior performance compared to the same network without the distillation technique. This superiority is further evident when juxtaposed with other baselines detailed in Table \ref{tab:example}.

Table \ref{tab:example} provides a summarized view of the decoding performance across various methods applied to EEG data. Clear trends in accuracy emerge across model types. Classical machine learning baselines, which utilize averaged or PCA-reduced EEG, yield near chance-level accuracy, underscoring the inadequacy of hand-engineered features for decoding intricate visual stimuli. An exception is the Logistic Regression model trained on squared data averages.

Conversely, deep learning models that harness spatiotemporal EEG TFDs patterns consistently achieve superior accuracy. Both convolutional and recurrent neural networks processing raw EEG time series deliver satisfactory results. Yet, the best performance is reached by models using 2D representations of multi-channel EEG. Specifically, CNNs fed with TFD computed using wavelet-transformed or spectrogram images both surpass $85\%$ in top-5 accuracy, underscoring the benefits of computer vision techniques that learn directly from 2D structures in signal processing. Both wavelet and spectrogram decompositions seem to encapsulate pertinent time-frequency domain information for decoding.

A closer examination of the top-3 and top-5 accuracy metrics reveals a consistent trend: deep learning models outclass classical baselines. The elite CNNs achieve over $75\%$ in top-3 accuracy, implying that in approximately 3 out of 4 trials, the true label ranks within the top three predictions. The performance gap relative to the LSTM network is also noteworthy. This accentuates the efficacy of 2D convolutions in discerning the pertinent semantic categories from EEG patterns. The consistency of the top-5 accuracy across deep learning models suggests potential inherent challenges in precisely mapping EEG to granular image labels. However, the models adeptly identify the overarching category within their top predictions, underscoring the viability of EEG-based visual concept decoding.

From a qualitative perspective, Figures \ref{fig:qualitative} and \ref{fig:other_qualitative} showcase examples of predicted and reconstructed images. While the model predominantly identifies the correct visual concept from EEG patterns, minor category confusions do arise. For instance, "bolete" might be misinterpreted as "pizza," or "banana" as "Margherita". Nevertheless, the model's ability to accurately discern the overarching semantic category and produce corresponding reconstructions is noteworthy.

In conclusion, our findings underscore the pivotal role of neural networks and image-centric representations in harnessing the rich multidimensional EEG representation. Directly classifying TFD inputs using a computer vision approach emerges as the potent strategy for EEG-based decoding.



\section{Discussion}

The primary objective of this study was to decode and reconstruct visual representations from EEG-recorded human brain activity. By employing deep convolutional neural networks trained on EEG TFD and guided by the CLIP-based knowledge distillation technique, we managed to predict image classes from the ImageNet dataset with an accuracy of $87\%$ in the top-5 category. This knowledge distillation approach yielded a marked improvement in performance when compared to a baseline model and other data processing methodologies. While the model's predictions were generally reliable for the majority of subjects, it did exhibit some confusion between closely related classes. The capability to extract the semantic content of image stimuli from non-invasive EEG recordings presents significant implications for the future of brain-computer interfaces. The methodology we developed for image reconstruction could potentially pave the way for a form of artificial vision, where decoded contents from a user's neural activity are visualized in real-time. Furthermore, our model introduces the possibility of innovative neurofeedback experiments, wherein subjects could receive instantaneous visual feedback of decoded EEG patterns, facilitating the voluntary self-regulation of brain states \cite{eeg_feedback}. However, the study is not without  limitations. EEG serves as a macroscopic lens into the brain's visual processing mechanisms. To address the limitations of EEG's spatial resolution, integrating it with other imaging techniques, such as fMRI, which boasts superior spatial resolution, is a promising avenue. Such multimodal strategies have shown potential in reconstructing images with a higher degree of detail \cite{ferrante2023brain,ozcelik_reconstruction_2022,ozcelik2023braindiffuser,Takagi2022.11.18.517004}. Also, the model in its current configuration has not been optimized for decoding images outside the 40 categories used in the experiment, suggesting a need for further refinement. The variability in EEG decoding abilities across different subjects or sessions, influenced by cognitive and neural factors, remains a topic that warrants deeper exploration. One of the significant concerns in EEG decoding revolves around the inadvertent extraction of personal perceptual data, which must be rigorously addressed. Our methodology places a strong emphasis on the creation of subject-specific models. This ensures that the decoding process is both consensual and uniquely tailored to the individual, mitigating potential ethical concerns. This approach not only necessitates voluntary participation but also minimizes the risk of misinterpretations due to the model's specificity to individual neural patterns. The rapid training methodology we have introduced also holds promise for real-time feedback paradigms using models tailored to individual subjects, with a couple of seconds in inference time needed to predict class and generate the image on an A100 GPU. As the field of deep learning and generative models continues to evolve, we anticipate parallel advancements in EEG decoding and reconstruction capabilities.

\section{Conclusions}

In this study, we demonstrated the potential of deep neural networks, coupled with generative diffusion models, to reconstruct visual experiences directly from non-invasive EEG recordings. The application of knowledge distillation from language-image pretraining enabled our convolutional decoder to effectively extract semantic information from brain activity patterns. This capability significantly surpassed the performance of classical signal processing baselines. By generating images based on the predicted labels, we were able to produce visualizations that closely align with the decoded neural activity. Our emphasis on creating subject-specific models not only ensures a certain degree of privacy but also underscores the unique capabilities of EEG data in decoding individual mental representations. These techniques, which focus on translating neural signals into their corresponding images, can kickstart significant advancements in the domains of brain-computer interfaces and neural prosthetics, as well as human-computer interaction research. Overall, our findings highlight the potential of non-invasive brain imaging as a tool to provide insights into the human cognitive experience.






\bibliographystyle{abbrv}
\bibliography{egbib}

\begin{thebibliography}{10}

\bibitem{bai2023dreamdiffusion}
Y.~Bai, X.~Wang, Y.~pei Cao, Y.~Ge, C.~Yuan, and Y.~Shan.
\newblock Dreamdiffusion: Generating high-quality images from brain eeg signals, 2023.

\bibitem{imagenet}
J.~Deng, W.~Dong, R.~Socher, L.-J. Li, K.~Li, and L.~Fei-Fei.
\newblock Imagenet: A large-scale hierarchical image database.
\newblock In {\em 2009 IEEE conference on computer vision and pattern recognition}, pages 248--255. Ieee, 2009.

\bibitem{eeg_feedback}
S.~Enriquez-Geppert, R.~J. Huster, and C.~S. Herrmann.
\newblock Eeg-neurofeedback as a tool to modulate cognition and behavior: A review tutorial.
\newblock {\em Frontiers in Human Neuroscience}, 11, 2017.

\bibitem{ferrante2023brain}
M.~Ferrante, F.~Ozcelik, T.~Boccato, R.~VanRullen, and N.~Toschi.
\newblock Brain captioning: Decoding human brain activity into images and text, 2023.

\bibitem{hinton2015distilling}
G.~Hinton, O.~Vinyals, and J.~Dean.
\newblock Distilling the knowledge in a neural network, 2015.

\bibitem{kavasidis_brain2image_2017}
I.~Kavasidis, S.~Palazzo, C.~Spampinato, D.~Giordano, and M.~Shah.
\newblock \textit{Brain2Image}: Converting brain signals into images.
\newblock In {\em Proceedings of the 25th {ACM} international conference on Multimedia}, pages 1809--1817. {ACM}.

\bibitem{Brain2Image}
I.~Kavasidis, S.~Palazzo, C.~Spampinato, D.~Giordano, and M.~Shah.
\newblock Brain2image: Converting brain signals into images.
\newblock In {\em Proceedings of the 25th ACM International Conference on Multimedia}, MM '17, page 1809–1817, New York, NY, USA, 2017. Association for Computing Machinery.

\bibitem{pywavelet}
G.~R. Lee, R.~Gommers, F.~Waselewski, K.~Wohlfahrt, and A.~O. Leary.
\newblock Pywavelets: A python package for wavelet analysis.
\newblock {\em Journal of Open Source Software}, 4(36):1237, 2019.

\bibitem{trainingtest}
R.~Li, J.~S. Johansen, H.~Ahmed, T.~V. Ilyevsky, R.~B. Wilbur, H.~M. Bharadwaj, and J.~M. Siskind.
\newblock Training on the test set? an analysis of spampinato et al. [31], 2018.

\bibitem{ozcelik_reconstruction_2022}
F.~Ozcelik, B.~Choksi, M.~Mozafari, L.~Reddy, and R.~VanRullen.
\newblock Reconstruction of {Perceived} {Images} from {fMRI} {Patterns} and {Semantic} {Brain} {Exploration} using {Instance}-{Conditioned} {GANs}, Feb. 2022.
\newblock arXiv:2202.12692 [cs, eess, q-bio].

\bibitem{ozcelik2023braindiffuser}
F.~Ozcelik and R.~VanRullen.
\newblock Brain-diffuser: Natural scene reconstruction from fmri signals using generative latent diffusion, 2023.

\bibitem{palazzo_generative_2017}
S.~Palazzo, C.~Spampinato, I.~Kavasidis, D.~Giordano, and M.~Shah.
\newblock Generative adversarial networks conditioned by brain signals.
\newblock In {\em 2017 {IEEE} International Conference on Computer Vision ({ICCV})}, pages 3430--3438. {IEEE}.

\bibitem{Palazzo}
S.~Palazzo, C.~Spampinato, I.~Kavasidis, D.~Giordano, and M.~Shah.
\newblock Generative adversarial networks conditioned by brain signals.
\newblock pages 3430--3438, 10 2017.

\bibitem{radford2021learning}
A.~Radford, J.~W. Kim, C.~Hallacy, A.~Ramesh, G.~Goh, S.~Agarwal, G.~Sastry, A.~Askell, P.~Mishkin, J.~Clark, G.~Krueger, and I.~Sutskever.
\newblock Learning transferable visual models from natural language supervision, 2021.

\bibitem{stable}
A.~Ramesh, P.~Dhariwal, A.~Nichol, C.~Chu, and M.~Chen.
\newblock Hierarchical text-conditional image generation with clip latents, 2022.

\bibitem{cebra}
S.~Schneider, J.~H. Lee, and M.~W. Mathis.
\newblock Learnable latent embeddings for joint behavioural and neural analysis.
\newblock {\em Nature}, 617(7960):360--368, May 2023.

\bibitem{singh_eeg2image_2023}
P.~Singh, P.~Pandey, K.~Miyapuram, and S.~Raman.
\newblock {EEG}2image: Image reconstruction from {EEG} brain signals.

\bibitem{spampinato2019deep}
C.~Spampinato, S.~Palazzo, I.~Kavasidis, D.~Giordano, M.~Shah, and N.~Souly.
\newblock Deep learning human mind for automated visual classification, 2019.

\bibitem{spampinato_deep_2017}
C.~Spampinato, S.~Palazzo, I.~Kavasidis, D.~Giordano, N.~Souly, and M.~Shah.
\newblock Deep learning human mind for automated visual classification.
\newblock In {\em 2017 {IEEE} Conference on Computer Vision and Pattern Recognition ({CVPR})}, pages 4503--4511. {IEEE}.

\bibitem{wavelet}
A.~Subasi.
\newblock Eeg signal classification using wavelet feature extraction and a mixture of expert model.
\newblock {\em Expert Systems with Applications}, 32(4):1084--1093, 2007.

\bibitem{Takagi2022.11.18.517004}
Y.~Takagi and S.~Nishimoto.
\newblock High-resolution image reconstruction with latent diffusion models from human brain activity.
\newblock {\em bioRxiv}, 2023.

\bibitem{zafar_decoding_2015}
R.~Zafar, A.~S. Malik, N.~Kamel, S.~C. Dass, J.~M. Abdullah, F.~Reza, and A.~H. Abdul~Karim.
\newblock Decoding of visual information from human brain activity: A review of {fMRI} and {EEG} studies.
\newblock 14(2):155--168.

\end{thebibliography}

\end{document}